\documentclass[twocolumn,showpacs,preprintnumbers,amsmath,amssymb]{revtex4}

\usepackage{graphicx}
\usepackage{dcolumn}
\usepackage{amsfonts,amsmath,amssymb,bm}

\begin{document}

\title{\bf {Polarizability of molecular chains: does one need exact exchange?}} 
\date{\today} 
\author{C.~D.~Pemmaraju$^a$}
\author{S. Sanvito$^a$}
\author{K. Burke$^b$}
\affiliation{a) School of Physics and CRANN, Trinity College, Dublin 2, Ireland\\
b) Departments of Chemistry and of Physics, UC Irvine, CA 92697}

\begin{abstract}

Standard density functional approximations greatly over-estimate the static polarizability of long-chain polymers,
but Hartree-Fock or exact exchange calculations do not.  Simple self-interaction corrected (SIC) approximations
can be even better than exact exchange, while their computational cost can scale only linearly with the number
of occupied orbitals.
\end{abstract}

\maketitle

Ground-state Kohn-Sham (KS) density functional theory (DFT) has become extraordinarily
popular for solving electronic structure problems in solid-state physics,
quantum chemistry and materials science \cite{FNM03}.  The accuracy of modern generalized
gradient approximations (GGAs) and hybrid functionals has proven sufficient for
many applications, often with surprisingly small errors.  Bond dissociation
energies, geometries, phonons, etc., are now routinely calculated with
errors of 10-20\%.

But local and gradient-corrected functionals 
overestimate massively the static polarizability and hyperpolarizability of molecular 
chains, especially conjugated polymers. This failure has been the 
subject of many studies over the last decade \cite{Champagne,Gisberg, Baerends1,Baerends2,Iikura,Abdurahman,Yang,
Kummel1,Kummel2}, studies which highlight the important role played by
 the response field originating from the exchange-correlation (XC) potential. 
The exact induced XC field counteracts the applied external field, keeping the polarization low.
In the local (or gradient-corrected) density approximation (LDA), this field erroneously points in the
{\em same} direction as the applied field \cite{Gisberg, Baerends1,Baerends2}.  Such failures of standard
functionals appear in other contexts, such as transport through single molecules \cite{Cormac}, or the
polarizability of large molecules.

In contrast, these effects are easily captured within standard wavefunction theory.  In particular,
Hartree-Fock (HF) theory does not greatly overestimate the polarizabilities and provides a good
starting point for more accurate wavefunction treatments, such as M\"oller-Plesset 
(MP) perturbation theory.
Thus exact exchange (EXX) DFT, the KS-DFT method for minimizing the HF energy
while retaining a single multiplicative potential, provides a promising alternative and indeed
has been found to give results very similar to HF \cite{Baerends1,Yang,Kummel2}.
This improvement can be attributed to the orbital-dependence of EXX, and the lack of
self-interaction error \cite{Yang}, i.e. EXX is exact for one electron, unlike LDA or GGA.

However, EXX is only one among many possible self-interaction free functionals
that one may construct. In fact any GGA can be corrected
to become self-interaction free (self-interaction corrected - SIC) by direct subtraction
of the XC functional evaluated on each of the individual orbitals \cite{pzsic}.  While
this can be performed for either LDA or GGA, only LDA has
significantly improved energetics from this procedure, but many investigators are
searching for useful methods to correct GGA's for self-interaction \cite{VSP06}. 
More importantly, EXX includes a sum over all {\em unoccupied} orbitals,
while SIC functionals use only occupied ones, i.e., EXX can often be substantially
more expensive computationally.
So the question then becomes: does one really need EXX,
or will any self-interaction free functional perform equally well~? 

We perform SIC calculations for the polarizabilities of hydrogenic chains using LDA and GGA.
Our SIC potential is constructed using the optimized
effective potential (OEP) framework within the Krieger-Li-Iafrate (KLI)
approximation \cite{KLI}. Using results from accurate wave-function methods as a benchmark, we find that
the polarizabilities calculated with KLI-SIC are in better agreement than those obtained with KLI exact exchange (X-KLI), 
with the remaining error attributed to the KLI approximation \cite{Kummel2,KummelCom}. This is an important result since KLI-SIC scales only
linearly with the number of occupied Kohn-Sham (KS) orbitals, as compared with the quadratic scaling of X-KLI. 
Thus SIC becomes a valuable scheme for evaluating the static polarizability of polymers with large unit
cells, and in other applications where these effects may be important \cite{Cormac}.

We start with a brief description of the SIC method used in this work. In DFT \cite{DFT}, the total energy functional 
$E[\rho^\uparrow,\rho^\downarrow]$ ($\rho^\sigma$ is the spin $\sigma=\uparrow, \downarrow$ density, 
$\rho=\sum_\sigma\rho^\sigma$) can be written as 
\begin{equation}
E[\rho^\uparrow,\rho^\downarrow]=T_\mathrm{S}[\rho]+\int\mathrm{d}^3{\bf r}\:\rho({\bf r})v({\bf r})+U[\rho]+E_\mathrm{xc}[\rho^\uparrow,\rho^\downarrow]\;,
\label{enfunc}
\end{equation}
with $T_\mathrm{S}$ the kinetic energy of the non-interacting KS orbitals, $v({\bf r})$ the external potential, $U$ the 
Hartree energy, and $E_\mathrm{xc}$ the XC energy.  For any GGA
\begin{equation}
E_\mathrm{xc}^\mathrm{SIC}[\lbrace\rho_n^\sigma\rbrace]=E_\mathrm{xc}^\mathrm{GGA}[\rho^\uparrow,\rho^\downarrow]-
\sum_{n\sigma}^\mathrm{occupied}(U[\rho_n^\sigma]+E_\mathrm{xc}^\mathrm{GGA}[\rho_n^\sigma,0])\;,
\label{EFSIC}
\end{equation}
where $\rho^\sigma_n=|\psi_n^\sigma|^2$ is the density of the $n$-th KS orbital.
Levy's minimization \cite{L82} leads to a set of single particle KS-like equations for $\psi_n^\sigma$ with 
corresponding eigenvalues $\epsilon_{n}^{\sigma,\mathrm{SIC}}$ and occupation numbers $p_n^\sigma$ 
($\rho^\sigma = \sum_{n} p_n^\sigma \rho_n^\sigma$)
\begin{equation}
\left[-\frac{1}{2}\boldsymbol{\nabla}^2+v_{\mathrm{eff},n}^\sigma({\bf r})\right]\psi_n^\sigma=\epsilon_{n}^{\sigma,\mathrm{SIC}}
\psi_n^\sigma\;.
\label{KSSIC}
\end{equation}

The effective potential $v_{\mathrm{eff},n}^{\sigma}({\bf{r}})$ is now KS-orbital dependent and cannot be classified 
as a standard multiplicative KS-potential. For instance it is ambiguously defined for unoccupied KS-orbitals since 
the SIC is only defined for the occupied ones. The solution of equation (\ref{KSSIC}) has followed several approaches.
The simplest one is to solve it directly under a normalization constraint with the resulting non-orthogonal orbitals 
undergoing an orthogonalization procedure \cite{pzsic}. This scheme however is not free of complications, since
$E_\mathrm{xc}^\mathrm{SIC}$ is not invariant under a unitary transformations of the occupied $\lbrace\psi_n^\sigma\rbrace$. 
The solution \cite{Harrison} is then to work with an auxiliary set of localized orbitals 
$\lbrace\phi_n^\sigma\rbrace$ used for constructing $v_{\mathrm{eff},n}^{\sigma}({\bf{r}})$ and related to $\lbrace\psi_n^\sigma\rbrace$ 
by a unitary transformation chosen so as to minimize $E_\mathrm{xc}^\mathrm{SIC}$.

A convenient alternative is offered by the OEP method \cite{Kummel3} where the orbital-dependent 
$v_{\mathrm{eff},n}^{\sigma}({\bf{r}})$ is recast into a local multiplicative orbital-independent potential. In this way
the SIC problem can then be solved as a normal KS problem. However the construction of an OEP is computationally
demanding. Here we adopt the KLI approximation \cite{KLI}, which is practically easy to construct and 
retains most of the advantages of the full OEP, being therefore well suited to the SIC problem \cite{Garza}.
The orbital-independent KLI effective SIC Kohn-Sham potential takes the form
\begin{multline}
v_\mathrm{KS}^{\sigma}(\mathbf{r}) = v(\mathbf{r})+ 
v_\mathrm{H}(\mathbf{r})+ v_\mathrm{xc}^{\sigma,\mathrm{GGA}}(\mathbf{r})+
v_\mathrm{xc}^{\sigma,\mathrm{SIC}}(\mathbf{r}),
\label{VKS}
\end{multline}
where $v_\mathrm{H}$ and $v_\mathrm{xc}^{\sigma,\mathrm{GGA}}$ are the Hartree and 
GGA-XC potentials.  We define the SIC potentials by
\begin{equation}
u_n^{\sigma,\mathrm{SIC}}(\mathbf{r})=
-v_\mathrm{Hxc}^{\sigma,\mathrm{GGA}}[\tilde{\rho}_n^{\sigma},0](\mathbf{r}),
\label{usic}
\end{equation}
where $v_\mathrm{Hxc}$ is the sum of the Hartree and GGA potentials, and their
Slater average as
\begin{equation}
w_\mathrm{xc}^{\sigma,\mathrm{SIC}}(\mathbf{r})=
\sum_{n=1}^{N^{\sigma}}\tilde{f}_{n}^{\sigma}(\mathbf{r})
u_n^{\sigma,\mathrm{SIC}}(\mathbf{r}),
\label{vslat}
\end{equation}
where $\tilde{f}_{n}^{\sigma}(\mathbf{r})=
\tilde{\rho}_{n}^{\sigma}(\mathbf{r})/\rho^{\sigma}(\mathbf{r})$
is the auxiliary orbital density as a fraction of the spin-density of its spin.
Then
\begin{equation}
v_\mathrm{xc}^{\sigma,\mathrm{SIC}}(\mathbf{r})=
w_\mathrm{xc}^{\sigma,\mathrm{SIC}}(\mathbf{r})+
\sum_{n=1}^{N^{\sigma}} {\tilde{f}_{n}^{\sigma}(\mathbf{r})
[\Delta v_n^{\sigma,\mathrm{SIC}} - C^{\sigma}]}.
\label{vxcsic}
\end{equation}
The orbital densities $\tilde{\rho}_n^{\sigma}$ in Eqs.~(\ref{usic})-(\ref{vxcsic}) are calculated from the 
auxiliary set of localized orbitals $\lbrace\phi_i^\sigma\rbrace$ instead of the canonical $\lbrace\psi_i^\sigma\rbrace$.  
Both sets of orbitals give the same total density $\rho^{\sigma}$. The orbital shift terms 
$\Delta v_n^{\sigma,\mathrm{SIC}}$ being a constant), are obtained by solving
\begin{multline}
\sum_{n=1}^{N^{\sigma}}
(\delta_{nm}-\overline{{\tilde{f}}}_{nm}^{\sigma})\Delta v_n^{\sigma,\mathrm{SIC}}=
\overline{w}_{\mathrm{xc},m}^{\sigma,\mathrm{SIC}}-
\overline{u}_{m}^{\sigma,\mathrm{SIC}}, 
\label{algeqn}
\end{multline}
where
$ \overline{h}_{m}^{\sigma}=
\int \mathbf{dr} \tilde{\rho}_m^{\sigma}(\mathbf{r})
h^{\sigma}(\mathbf{r})$.
Furthermore, 
\begin{equation}
\sum_{n=1}^{N^{\sigma}} \overline{{\tilde{f}}}_{nm}^{\sigma}=1\:,\;\;\;\;\;\;  
\sum_{n=1}^{N^{\sigma}} (\overline{w}_{\mathrm{xc},m}^{\sigma,\mathrm{SIC}}-
\overline{u}_{m}^{\sigma,\mathrm{SIC}}) = 0 \:,
\end{equation}
so that the linear system in Eq.(\ref{algeqn}) is of rank ($N^{\sigma}$-1) and must be solved using a least 
squares approach. The constant $C^{\sigma}$ is set to $\Delta v_\mathrm{HOMO}^{\sigma,\mathrm{SIC}}$
(HOMO is the highest occupied molecular orbital). We have implemented this scheme in the DFT code SIESTA \cite{siesta}.

To facilitate easy comparison with previous quantum chemistry and EXX-DFT calculations \cite{Kummel2}, we chose as a test
system the widely studied linear hydrogen chains H$_n$ made up of $n$ H atoms with alternating H-H distances of 2 and 3 
$a_0$. An optimized atomic orbital basis set consisting of double zeta polarized $s$ and triple zeta polarized $p$
functions is employed.
\begin{figure}[htb]
\includegraphics[width=0.4\textwidth,clip=true]{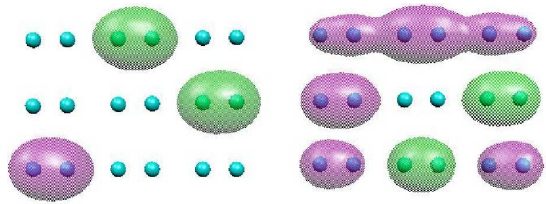}
\caption{\label{lmo}The Pipek-Mezey localized molecular orbitals $\lbrace\phi_i^\sigma\rbrace$ (left) and 
canonical Kohn-Sham orbitals $\lbrace\psi_i^\sigma\rbrace$ (right) for the occupied states of the H$_\mathrm{6}$ molecule.}
\end{figure}
The Pipek-Mezey localization scheme \cite{pipek} which minimizes the number of atoms over which a given molecular 
orbital is delocalized, is used to transform the canonical KS-orbitals $\lbrace\psi_n^\sigma\rbrace$
into the localized set $\lbrace\phi_n^\sigma\rbrace$. As an example both the $\lbrace\phi_n^\sigma\rbrace$ and the
$\lbrace\psi_n^\sigma\rbrace$ of H$_6$ are presented in figure \ref{lmo}. Finally the static polarizability 
$\alpha = \delta\mu_{z}/\delta F_{z}$ is calculated numerically using finite differences. 

\begin{table*}[htb]
\caption{\label{Tab1}Calculated polarizability $\alpha$ of H$_n$ chains obtained by using the KLI-SIC method 
with different XC functionals. The subscript X indicates that correlation has been dropped from the XC potential.
These are compared with MP4, HF and exact exchange DFT (EXX) results from reference \cite{Kummel2}. }
\begin{ruledtabular}
\begin{tabular}{|l|c|cc|ccc|ccc|cccc|}
&exact &\multicolumn{2}{c|}{GGA} &\multicolumn{3}{c|}{EXX} &\multicolumn{3}{c|}{KLI-SIC} &\multicolumn{4}{c}{ X-only}\\
H$_{N}$ &MP4 &LDA &PBE &HF &X-OEP &X-KLI &LDA &PBE &LDA-OEP\footnote{Estimated by subtracting the difference between
X-KLI and X-OEP.}\cite{KummelCom} &LDA$_\mathrm{X}$ &PBE$_\mathrm{X}$ &SIC-LDA$_\mathrm{X}$&SIC-PBE$_\mathrm{X}$\\
\hline
H$_4$   &29.5   &37.26   &35.62   &32.0   &32.2   &33.11   &33.38    &33.14  &32.5  &38.90   &36.51   &33.37&33.10\\
H$_6$   &51.6   &73.10   &69.35   &56.4   &56.6   &60.64   &58.56    &58.07  &54.6 &76.16   &70.45   &58.84&57.63\\
H$_8$   &75.9   &116.58  &109.74  &82.3   &84.2   &91.56   &86.94    &86.48  &76.4 &121.64  &110.84  &87.08&84.53\\
H$_{10}$&--     &166.36  &155.31  &--     &--     &124.87  &117.28   &116.16 &--   &173.89  &156.45  &116.77&113.14\\
H$_{12}$&126.9  &220.55  &204.53  &137.6  &138.1  &159.27  &147.96   &145.98 &126.8 &231.25  &205.51  &147.19&141.90\\
\end{tabular}
\end{ruledtabular}
\end{table*}
Table \ref{Tab1} contains the central results of this work.
The first column contains highly accurate (MP4) quantum chemical results, which
we take as exact.  The next two columns show LDA and PBE results \cite{pbe}, 
demonstrating the LDA overestimate (by about 100\% for H$_{12}$) of $\alpha$, an
overestimate that is only slightly reduced by GGA.  We checked several GGA's,
and they all had the same features. Small finite systems, such as atoms, are well-known to overpolarize
in LDA and GGA, because their XC potentials are too shallow, and do
not decay with the correct asymptotic form, $-1/r$.   But this is
a distinct effect from that discussed here, which is due to the response
to the electric field inside the molecule. Ours is a bulk effect, not
depending on the end points. We checked this and found that the LB94 functional \cite{lb94}, specifically designed 
to reproduce the correct asymptotic
behavior, does not yield any better results than the other GGA's and in fact worsens the LDA $\alpha$.

The next 3 columns list results of different types of calculations using the Fock integral, and no
correlation.  We note that HF slightly overestimates the polarizability, but by less than 10\%.
An exact OEP treatment of the same functional (X-OEP) yields essentially the same numbers.  Our KLI
scheme, applied to the same functional (X-KLI), makes a noticeable overestimate, but the error remains
less than 20\%, compared to OEP.

Now we focus on the SIC results. Regardless of the GGA functional used, the KLI-SIC
polarizabilities show a drastic improvement with respect to those obtained with pure GGA.  Furthermore,
PBE results are very close to LDA results.  There remains about a 20\% overestimate for H$_{12}$, but this
is probably due to the KLI approximation. Correcting the KLI-SIC-LDA values, using the difference between KLI and OEP for EXX as an error estimate, suggests that OEP-SIC results(\textit{Cf.} Korzdorfer \textit{et al.}\cite{KummelCom}) might be in good agreement with MP4. These corrected values are listed under LDA-OEP in Table \ref{Tab1}.

We also checked if the inclusion of correlation was important for the excellent KLI-SIC results,
by running the calculations with correlation removed.  For the pure functionals, LDA correlation slightly
reduces the huge overestimate, whereas PBE correlation has almost no effect.  On the other hand, for
the KLI-SIC results, LDA correlation has almost no effect, while PBE correlation corrects the PBE$_\mathrm{X}$ 
result in the {\em wrong} direction.  This is consistent with the general result that SIC-GGA includes some
incorrect overcounting of gradient effects \cite{VSP06}.

\begin{figure}[htb]
\includegraphics[width=0.5\textwidth,clip=true]{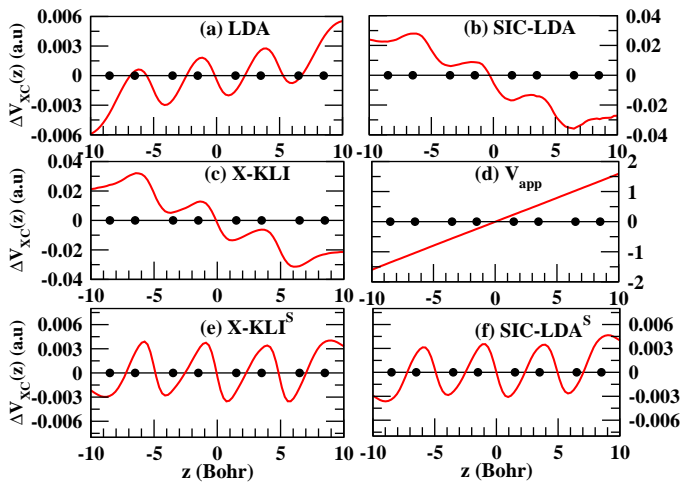}
\caption{\label{xcresp}$\Delta V_\mathrm{XC}(z)=V_\mathrm{XC}^\mathrm{E}(z)-V_\mathrm{XC}$ for H$_8$ plotted 
against position ($\mathrm{z}$) for (a) LDA, (b) SIC-LDA, (c) X-KLI, (e) X-KLI$^\mathrm{S}$ and (f) SIC-LDA$^\mathrm{S}$. 
The applied external field is shown in (d) and the black dots indicate the position of the H atoms. }
\end{figure}

\begin{table}[htb]
\caption{\label{Tab3}Static polarizability $\alpha$ of H$_n$ obtained from SIC and X-KLI where only the Slater 
term in the KLI potential is used (super-script S).  Both XC and X-only results are shown for the SIC-LDA$^\mathrm{S}$ 
and SIC-PBE$^\mathrm{S}$.}
\begin{ruledtabular}
\begin{tabular}{lccccc}
H$_{N}$&SIC-LDA$^\mathrm{S}$&SIC-PBE$^\mathrm{S}$&SIC-LDA$^\mathrm{S}_\mathrm{X}$&SIC-PBE$^\mathrm{S}_\mathrm{X}$&X-KLI$^\mathrm{S}$\\
\hline
H$_4$&35.37&35.12&36.25&35.16&35.78\\
H$_6$&67.74&67.13&68.92&66.34&69.17\\
H$_8$&105.91&104.89&107.57&102.97&108.72\\
H$_{10}$&148.64&146.18&150.86&143.10&152.90\\
H$_{12}$&193.94&190.47&197.05&185.25&199.91\\
\end{tabular}
\end{ruledtabular}
\end{table}
The improved response obtained with orbital dependent functionals is due the opposing XC field, already demonstrated in the literature for OEP exact exchange\cite{Kummel2}.  Here we verify that the same happens with the KLI-SIC scheme. In 
figure \ref{xcresp} we plot $\Delta V_\mathrm{XC}$ defined as the difference between the XC potential with 
and without an applied electric field $\Delta V_\mathrm{XC}(z)=V_\mathrm{XC}^\mathrm{E}(z)-V_\mathrm{XC}$ 
for LDA, SIC-LDA and X-KLI.   We denote by a superscript S results obtained by including only the Slater average 
potential in Eq. (\ref{vxcsic}), dropping the constant terms.  Clearly both contributions are significant in the final 
XC potential. The Slater-only polarizabilities are shown in table \ref{Tab3} and reflect this.

\begin{figure}[b]
\includegraphics[width=0.3\textwidth,clip=true]{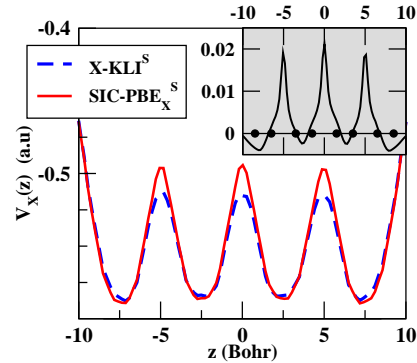}
\caption{\label{pslat} XC potential plotted against distance ($\mathrm{z}$) for SIC-PBE$^\mathrm{S}$ (solid line) and 
X-KLI$^\mathrm{S}$. The grey inset plots the difference between the two potentials.}
\end{figure}

By comparing the values of $\alpha$ in the tables \ref{Tab1} and \ref{Tab3} one may conclude that
the Slater average already contains important corrections, but the bulk of the effect is contained in
the orbital shift term. For example if one considers the $\alpha$ calculated with LDA for H$_{12}$
the polarizability is 220.5 for LDA, 193.9 for SIC-LDA$^\mathrm{S}$ and 147.9 for SIC-LDA. It is also
interesting to note that even at the level of the Slater average, SIC performs better than X-KLI. 
This can be understood by looking at the XC potential for SIC-PBE$^\mathrm{S}$ and 
X-KLI$^\mathrm{S}$ (Fig.~\ref{pslat}) when no external field is applied. 
The SIC-PBE potential exhibits higher peaks in the inter-molecular space between 
H$_\mathrm{2}$ units than X-KLI. This explains the quantitative difference in $\alpha$ between the 
two cases. The improved performance of X-OEP over X-KLI can be attributed \cite{Kummel2} 
to similar barriers in the inter-molecular region which however arise from the response part of the X-OEP potential.

Finally we ask whether similar results can be obtained with atomic-like corrections, which have the effect
of making the KS potential deeper at the atomic sites.
We investigated both the LDA+U \cite{ldau} and the atomic LDA-SIC (ASIC)
methods \cite{Alessio,ASIC} in this regard.  The LDA+U results are extremely poor, as
it provides polarizabilities larger than even those obtained with simple LDA.
The ASIC results are far more promising, being half-way
between those of GGA and of SIC ($\alpha$=33.95, 63.67, 98.39, 137.42, 178.87 for H$_n$ respectively with $n$=
4, 6, 8, 10, 12). In fact, if the ASIC corrections relative to pure LDA were double those we found, they would reproduce 
the HF results very accurately.

The above can be understood by the way the atomic corrections are introduced.
The H atoms are half-filled in the absence of an external field. 
Switching on the field produces a tiny charge transfer, which however is amplified by the LDA+U potential. As a result the already wrong LDA response field gets amplified. The same does not happen with ASIC, which improves the response over LDA by virtue of 
the higher inter-molecular barriers.

In conclusion we have demonstrated that SIC functionals at the KLI level, in general, perform
better than X-KLI at computing the static polarizability of hydrogenic chains. This is an interesting result
in view of the considerably less computational overheads involved with the SIC method with respect to 
EXX. Our work therefore opens the prospect of using SIC for evaluating the electrical response
of complex polymeric materials. 

This work is funded by the Science Foundation of Ireland (SFI02/IN1/I175)
and by the US Department of Energy (DE-FG02-01ER45928).

\end{document}